\documentclass[prl,twocolumn,showpacs,amsmath,amssymb,citeautoscript,nobibnotes,preprintnumbers]{revtex4-1}

\usepackage{graphicx,times}
\usepackage{array}
\usepackage{bm}
\usepackage{amsfonts}
\usepackage{graphics}
\usepackage{mathrsfs}
\usepackage{amssymb}
\usepackage{amsmath}
\usepackage[utf8]{inputenc}
\usepackage[T1]{fontenc}
\usepackage{parskip}
\usepackage{float}

\newcommand{\beq}{\begin{equation}}
\newcommand{\eeq}{\end{equation}}
\newcommand{\beqa}{\begin{eqnarray}}
\newcommand{\eeqa}{\end{eqnarray}}

\newcommand{\psib}{{\overline{\psi}}}

\newcommand\comment[1]{}

\begin{document}

\title{Solution to sign problems in half-filled spin-polarized electronic systems}

\author{Emilie Fulton Huffman and Shailesh Chandrasekharan}
\affiliation{Department of Physics, Duke University, Durham, NC 27708, USA}

\keywords{Sign Problem, Four-Fermion models, Half-Filling}
\begin{abstract}
We solve the sign problem in a particle-hole symmetric spin-polarized fermion model on bipartite lattices using the idea of fermion bags. The solution can be extended to a class of models at half filling but without particle-hole symmetry. Attractive Hubbard models with an odd number of fermion species can also be solved. The new solutions should allow us to study quantum phase transitions that have remained unexplored so far due to sign problems.
\end{abstract}
\pacs{71.10.Fd, 71.27.+a,71.30.+h}
\maketitle

Quantum Monte Carlo methods for many body fermionic systems in thermal equilibrium usually require one to be able to rewrite quantum partition functions as a sum over classical configurations with positive Boltzmann weights that are computable in polynomial time. Unfortunately, due to the underlying quantum nature of the problem, the Boltzmann weights can be negative or even complex in general. Such expansions are said to suffer from a sign problem since they cannot be used in Monte Carlo sampling \cite{PhysRevB.41.9301}. The discovery of an expansion with positive Boltzmann weights is referred to as a solution to the sign problem. Solutions to sign problems in many quantum systems are considered to be outstanding problems in computational complexity \cite{PhysRevLett.94.170201}.

Traditionally, solutions are based on rewriting the interacting problem as a free fermion problem where fermions only interact with background auxiliary fields \cite{Fucito:1980fh,PhysRevLett.46.519,PhysRevD.24.2278,PhysRevLett.47.1628}. The Boltzmann weight then depends on the determinant of the free fermion matrix, which can still be negative or complex. However, in electronic systems, a symmetric treatment of both spin components of the electron can sometimes make the Boltzmann weight positive since it can be written as the product of two real determinants that come with the same sign \cite{PhysRevLett.56.2521}. Sign problems in spin-polarized systems are usually much harder to solve since the Boltzmann weight contains only a single determinant. In certain cases the presence of an anti-unitary symmetry in the fermion matrix can help prove the absence of sign problems even though there is only a single fermion determinant \cite{Hands:2000ei}. However, such an approach also usually requires the presence of an even number of fermion species.

Spin-polarized electronic systems with particle-hole symmetry are special since holes can mimic the second species of fermions. In relativistic systems, particle-hole symmetry is replaced by charge conjugation symmetry and anti-particles can play the role of the second species of fermions. Thus one might expect that solutions to sign problems would emerge naturally in the presence of particle-hole or charge conjugation symmetries. Indeed, many difficult sign problems do arise away from half filling where these symmetries are broken. However, even in the presence of these symmetries, solutions to sign problems typically seem to require a second species of fermions. It is easy to find models of spin-polarized electronic systems with particle-hole symmetry and single flavor relativistic fermion models with charge conjugation symmetry that suffer from sign problems in the traditional approach. 

Consider for example the tight binding model of spin-polarized graphene where the repulsion between the electrons is modeled with a nearest neighbor Hubbard-type interaction. The Hamilton operator is given by
\begin{equation}
H = \sum_{\langle ij\rangle} -t (c^\dagger_i c_j + c^\dagger_j c_i) + 
V\Big(n_i - \frac{1}{2}\Big)\ \Big(n_j - \frac{1}{2}\Big),
\label{model}
\end{equation}
where $\langle ij \rangle$ refers to the nearest neighbor bond connecting different sublattices on the honeycomb lattice. The model is well known as the $tV$-model and was considered on square lattices a long time ago \cite{PhysRevB.29.5253,PhysRevB.32.103}. Although the model has a particle-hole symmetry, as far as we know its sign problem has not been solved by traditional methods for any value of $V$. It is useful to remember that with a single spin component, $V < 0$ cannot be mapped into $V > 0$ through a unitary transformation. Such a mapping usually requires two spin components and a restricted type of interaction. In the repulsive case for $V \geq 2t$ the sign problem could be solved using a non-traditional method called the meron-cluster approach \cite{PhysRevLett.83.3116}. Unfortunately, that solution could not be extended to smaller values of $V$ where there is an interesting quantum phase transition.

The above model is a regularized version of a massless four-fermion quantum field theory containing a single flavor of four-component Dirac fermions \cite{Rosenstein:1988pt,Hands:1992be}. Similar physics can also be obtained from a Hamiltonian description of staggered fermions on a square lattice \cite{PhysRevD.16.3031,Chandrasekharan:1999ys}. Massless four-fermion field theories help in describing quantum phase transitions from a semi-metal phase (containing massless Dirac fermions) to a Mott insulating phase (with massive Dirac fermions) accompanied by spontaneous symmetry breaking \cite{PhysRevB.80.075432}. Most quantum Monte Carlo studies of this phase transition involve an even number of flavors of four component Dirac fermions \cite{Christofi:2007ye,Chandrasekharan:2011mn,PhysRevX.3.031010}, while theories with an odd number of flavors remain unexplored due to sign problems. Predictions for the associated critical exponents using various analytic techniques do exist for a variety of four-fermion field theories including odd flavor numbers \cite{Rosa:2000ju,PhysRevLett.97.146401,Janssen:2012pq}.  

In this paper we solve the sign problem in (\ref{model}) for all values of $V > 0$, thus allowing us to study the quantum phase transition in it for the first time. While most of our discussion will be focused on (\ref{model}) for concreteness, many of the ideas behind the solution are general and easily extendable to other models including those with an odd number of fermion flavors. We will mention some of these extensions towards the end. 

We first rewrite the Hamilton operator (\ref{model}) in a form that makes particle-hole symmetry more explicit. Hence we write
\begin{equation}
H = \sum_{i,j} c^\dagger_i M_{ij} c_j + \frac{V}{4} \big(n^+_i - n^-_i) (n^+_j - n^-_j)
\label{model1}
\end{equation}
where $M$ is chosen appropriately, $n^+_i = n_i = c^\dagger_i c_i$ represents the particle number operator, and $n^-_i = (1-n_i) = c_i c^\dagger_i$ the hole number operator at site $i$. We call the free term $H_0$ and the interaction term
\begin{equation}
H_{\rm int} = \frac{V}{4}\ \sum_{b,s_i,s_j} (s_i n^{s_i}_i) \ (s_j n^{s_j}_j).
\label{intchoice}
\end{equation}
Here $b = \langle ij\rangle$ labels the bond connecting the nearest neighbor sites $i$ and $j$, and $\{s_i,s_j\} = \pm 1$ label the presence of either $n^+$ or $-n^-$ at the sites $i$ and $j$. The free matrix $M$ has the special property that it is real and only connects sites on opposite sublattices. This implies that
\begin{equation}
M^T = -D M D
\label{symm}
\end{equation}
where $D_{ij} = \sigma_i \delta_{ij}$ is a diagonal matrix with elements $\sigma_i = +1$ if $i$ belongs to the even sublattice and $\sigma_i = -1$ if $i$ belongs to the odd sublattice. This property of $M$ will play an important role in the solution to the sign problem.

Instead of the traditional auxiliary field method, we use the well known series expansion of the partition function that is used often these days in continuous time Monte Carlo methods \cite{PhysRevLett.81.2514,PhysRevB.72.035122,PhysRevE.74.036701,PhysRevLett.101.090402,PhysRevA.82.053621,RevModPhys.83.349}. The expansion is in powers of interaction vertices and in our model we obtain
\begin{eqnarray}
Z \ &=& \ Z_0 \sum_k \sum_{[b,s]}\ \int [dt] \big(-V/4\big)^k\ \nonumber \\
&& \mathrm{Tr}\Big(\mathrm{e}^{-(\beta-t_1) H_0} (s_1 n^{s_1}_{i_1}) \mathrm{e}^{-(t_2-t_1)H_0} (s_2 n^{s_2}_{i_2})\  ...
\nonumber \\
&& \hskip0.5in ...\ \mathrm{e}^{-(t_{2k-1}-t_{2k}) H_0} (s_{2k} n^{s_{2k}}_{i_{2k}}) \mathrm{e}^{-t_k H_0} \Big).
\label{sse}
\end{eqnarray}
where $Z_0$ is the free partition function, $[b,s]$ defines a configuration of $k$ interaction bonds located at times $t_1 \geq t_3 \geq t_5,\geq...\geq t_{2k-1}$, and $[dt]$ represents the $k$ time-ordered integrations from $0$ to $\beta$ over these locations of the interaction bonds. Each of the $k$ interaction bonds contains two interaction vertices. We label these $2k$ interaction vertices with the index $q=1,2...,2k$ such that $i_q$ labels the spatial site of the vertex, $t_q$ labels the temporal location of the vertex, and $s_q$ labels the particle-hole operator that is inserted at the vertex. Further since the two interaction vertices on each both occur at the same time, we naturally have $t_1 = t_2 \geq t_3 = t_4 \geq ...\geq t_{2k-1}= t_{2k}$. Thus the integration $[dt]$ involves only $k$ integrations as explained above. 

Using standard manipulations in the Fock space formalism, the trace in (\ref{sse}) can be evaluated exactly in terms of a determinant of a $2k \times 2k$ matrix $G([b,s,t])$ allowing us to write
\begin{equation}
Z \ = \ Z_0 \sum_k \sum_{[b,s]}\ \int\ [dt]\ (-V/4)^k \  \mathrm{Det} G([b,s,t]).
\label{sse1}
\end{equation}
The matrix elements of $G([b,s,t])$ can be obtained from the two point functions
\begin{subequations}
\begin{eqnarray}
\mathrm{Tr}\Big(\mathrm{e}^{-(\beta-t)H_0} c_{i_q}  \mathrm{e}^{-t H_0} c^\dagger_{i_{q'}}\Big) &=& \Big(\frac{\mathrm{e}^{-tM}}{1 + \mathrm{e}^{-\beta M}}\Big)_{i_q,i_{q'}} \\
\mathrm{Tr}\Big(\mathrm{e}^{-(\beta-t)H_0} c^\dagger_{i_q}  \mathrm{e}^{-t H_0} c_{i_{q'}}\Big) &=& \Big(\frac{\mathrm{e}^{t M^T}}{1 + \mathrm{e}^{\beta M^T}}\Big)_{i_q,i_{q'}}
\end{eqnarray}
\end{subequations}
using the well known Wick's theorem. We can then use (\ref{symm}) to write
\begin{equation}
\Big(\frac{\mathrm{e}^{t M^T}}{1 + \mathrm{e}^{\beta M^T}}\Big)_{i_q,i_{q'}} = \ \ 
\sigma_{i_q} \sigma_{i_{q'}} \Big(\frac{\mathrm{e}^{-t M}}{1 + \mathrm{e}^{-\beta M}}\Big)_{i_q,i_{q'}},
\end{equation}
where $\sigma_{i_q}$ are the diagonal elements of the matrix $D$. Using these results it is possible to prove that for a fixed $q < q'$ the off-diagonal matrix elements of $G([b,s,t])$ are given by
\begin{subequations}
\begin{eqnarray}
G_{qq'}([b,s,t]) \ &=&\ \Big(\frac{\mathrm{e}^{-(t_q-t_{q'}) M}}{1 + \mathrm{e}^{-\beta M}}\Big)_{i_q,i_{q'}},
\label{offdiag1}
\\
G_{q'q}([b,s,t]) \ &=&\ -\ \sigma_{i_q} \ \sigma_{i_{q'}}\ G_{qq'}([b,s,t]).
\label{offdiag2}
\end{eqnarray}
\label{offdiag}
\end{subequations}
The negative sign in (\ref{offdiag2}) is due to the usual anti-periodic boundary conditions in time that must be introduced when the trace is written as a determinant. Note that the off-diagnoal matrix elements depend on $[b,t]$ but not on $[s]$. For $q = q'$ we obtain the diagonal matrix elements which are given by 
\begin{equation}
G_{qq'}([b,s,t]) \ =\ \frac{s_q}{2}\ \delta_{q q'}
\label{diag}
\end{equation} 
and depend only on $[s]$ and not on $[b,t]$. It may be surprising that the variable $[s]$ does not enter the off-diagonal matrix elements (\ref{offdiag}). However, note that the insertion of $s_i n^{s_i}_i$ implies either $c^\dagger_i c_i$ or $-c_i c^\dagger_i$, both of which are the same operator except for diagonal terms. Hence it is natural that $[s]$ only enters through diagonal terms.

\begin{figure*}[t]
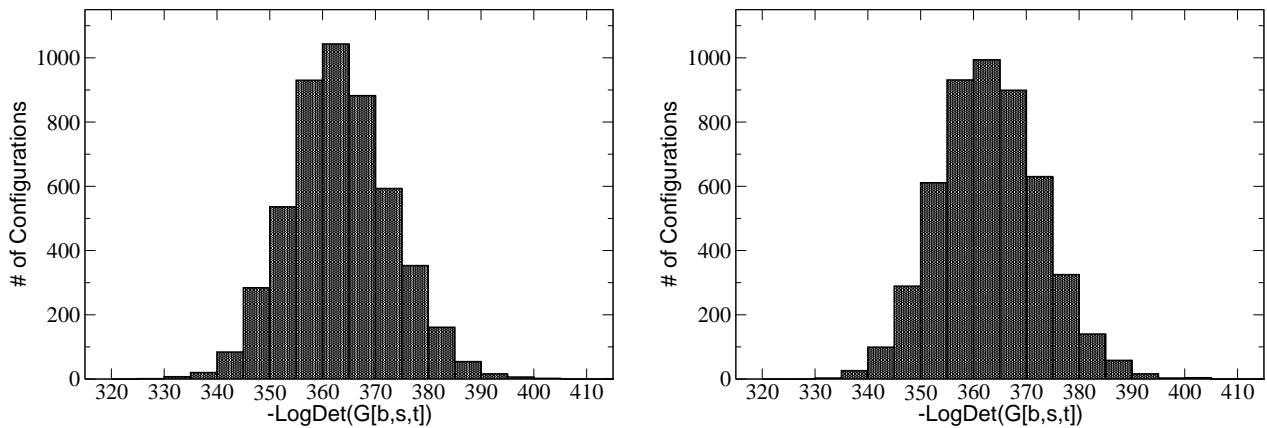

\begin{center}
\hbox{
\includegraphics[width=0.45\textwidth]{pos.eps}
\hskip0.2in
\includegraphics[width=0.45\textwidth]{neg.eps}
}
\caption{ In a sample of 10000 randomly generated configurations $[b,s,t]$ on an $8 \times 8$ square lattice at $\beta=10$ involving the matrices of size  $250 \times 250$, we obtained $4972$ positive determinants and $5028$ negative determinants. A histograms of positive determinants (left) and negative determinants (right) in this sample is plotted above. The similarity of the two distributions suggests the existence of a severe sign problem.
}
\label{fig1}
\end{center}
\end{figure*}

In a class of models like the attractive Hubbard model with two spin components, the trace is equal to the square of a determinant, one form each spin component. The Boltzmann weight is then positive and there is no sign problem allowing one to develop Monte Carlo algorithms. This approach is referred to as the diagrammatic determinantal Monte Carlo method which has been used to uncover the physics of the BCS-BEC cross-over recently \cite{PhysRevLett.101.090402,PhysRevA.82.053621}. In the current situation, unfortunately, there is no good reason for $\mathrm{Det}G([b,s,t])$ in (\ref{sse1}) to be positive and hence the approach does not naturally solve the sign problem. In order to demonstrate the existence of a sign problem, we have studied the behavior of $\mathrm{Det} G([b,s,t])$ on a two dimensional periodic square lattice of length $L=8$ at $\beta = 10$. We generated $10000$ random $[b,s,t]$ configurations containing $k=125$ interaction bonds (or equivalently $250$ interaction vertices). We found $4972$ configurations with positive determinants and $5028$ configurations with negative determinants. Figure \ref{fig1} shows the distribution of positive and negative determinants. As expected the similarity of the two distributions suggests a severe sign problem.

In order to solve the sign problem, we use the fermion bag approach introduced recently by one of us \cite{PhysRevD.82.025007}. The basic idea is to recognize that determinants help sum over fermion worldlines. Instead of summing over all possible fermion worldlines as is usually done, it may be possible to identify regions of space-time (called fermion bags) where the sum over fermion worldlines along with other sums of the partition function may naturally lead to positive weights. In other words it possible to reorganize the sums of the partition function more cleverly. A variety of unsolved sign problems have been solved using this approach \cite{PhysRevD.85.091502,PhysRevD.86.021701}. A review of the method along with examples of solvable models can be found in \cite{Chandrasekharan:2013rpa}. Here we show that (\ref{model}) is yet another example of a solvable model. As explained in \cite{Chandrasekharan:2013rpa}, diagonal elements of the fermion matrix can be treated as fermion bags. Since the variables $[s]$ appear only through diagonal terms, a sum over all possible $[s]$ configurations should be possible by treating these diagonal terms as fermion bags. In the current problem since the variable $s$ multiplies the diagonal term a sum over $s$ will lead to a zero weight. This means diagonal terms will not contribute to the partition function and can be set to zero. While this fact is intuitively clear from the fermion bag approach, it can also be easily derived mathematically using Grassmann variables. For example let us write $G([b,s,t]) = D_0([s]) + A([b,t])$ where $D_0([s])$ is the diagonal part defined in (\ref{diag}) and $A([b,t])$ is the off diagonal part defined in (\ref{offdiag}). Using Grassmann integrals we can write
\begin{eqnarray}
\sum_{[s]} \mathrm{Det}(G[b,s,t]) &=& \sum_{[s]} \int [d\psib \ d\psi] \mathrm{e}^{-\psib (D_0([s]) + A([b,t]))\psi} \nonumber \\
\end{eqnarray}
Substituting $D_0([s])$ from (\ref{diag}) and performing the $[s]$ sum first we obtain
\begin{equation}
\sum_{[s]} \ \mathrm{e}^{-\psi D_0([s])\psi} = \prod_q \sum_{s_q=1,-1} (1 - \frac{s_q}{2} \psib_q \psi_q) = 4^k,
\end{equation}
which means
\begin{equation}
\sum_{[s]} \mathrm{Det}(G[b,s,t]) = 4^k \mathrm{Det} (A([b,t])).
\end{equation} 
Substituting this result into (\ref{sse1}) we obtain
\begin{equation}
Z \ = \ \sum_{[b]}\ \int\ [dt]\ (-V)^k \ \mathrm{Det} (A([b,t])),
\label{sse2}
\end{equation}
where we have performed the sum over all $[s]$ configurations. As expected the diagonal terms disappear from $Z$. We will argue below that there is no sign problem in (\ref{sse2}).

The matrix $A([b,t])$ defined in (\ref{offdiag}) is real and satisfies the relation $A^T = - \tilde{D} A \tilde{D}$\textit{,} where $\tilde{D}$ is the diagonal matrix obtained from $D$ but restricted to the $2k$ interaction sites (i.e., $(\tilde{D})_{i_q,i_{q'}} = \sigma_{i_q}\ \delta_{i_q,i_{q'}}$). Hence, $(A\tilde{D})$ is a real anti-symmetric matrix whose determinant must be positive.  But $\mathrm{Det}(\tilde{D}) = (-1)^k$ since $k$ sites belong to the even sublattice and $k$ sites belong to the odd sublattice. Thus,
\begin{equation}
(-1)^k\mathrm{Det}(A([b,t]))\ =\ \mathrm{Det}(A([b,t])\tilde{D})  \geq 0
\end{equation}
Thus we finally obtain
\begin{equation}
Z \ = \ \sum_{[b]}\ \int\ [dt]\ (V)^k \  \mathrm{Det} (A([b,t])\tilde{D})
\end{equation}
which is a sum over positive terms for $V > 0$.

The above solution can easily be extended to other models, including models without particle-hole symmetry. To appreciate this, let us review the rules that models should satisfy in order to be free of sign problems
by the above approach. First, the free term can be modified as long as (\ref{symm}) can be maintained and $A\tilde{D}$ remains real and symmetric. It is easy to verify that the Hamilton operator for free staggered fermions satisfies these constraints, allowing us to explore a single four-component Dirac fermions in three dimensions\cite{Chandrasekharan:1999ys}. As far as we know this has not been possible so far. Secondly, we define a {\em staggered reference configuration} which contains a particle on the even sublattice and a hole on the odd sublattice. Interaction vertices that respect this reference configuration do not cause sign problems. This means interaction vertices may contain a particle number operator $n^+_i$ on the even sublattice and/or a hole number operator $n_i^-$ on the odd sublattice. If a vertex violates the reference configuration then sign problems will in general be introduced. However, if violations can be introduced in a correlated fashion such that a controlled resummation over positive and negative configurations can be performed, then sign problems can again be solved. In the fermion bag language this means a clever choice of the fermion bags may allow for the resummation and lead to positive weights. Couplings of the type $V$ in (\ref{model1}) are examples of such correlated couplings that violate the reference configuration yet do not cause sign problems. Based on these rules we see that the nearest neighbor interaction can be generalized to the form
\begin{equation}
H_{\rm int} = \sum_{i,j} V_{ij} \Big(n_i - \frac{1}{2}\Big) \Big(n_j - \frac{1}{2})
\end{equation}
where $V_{ij} \geq 0$ when $i$ and $j$ belong to opposite sublattice and $V_{ij} \leq 0$ when they belong to the same sublattice. It is also possible to introduce a staggered chemical potential term,
\begin{equation}
H_{\rm stagg} = \sum_i h_i n^{s_i}_i
\end{equation}
where $h_i \geq 0$ and $s_i$ is $+1$ on the even sublattice and $-1$ on the odd sublattice.

In addition to spin-polarized models, our solution extends easily to models with an odd number of interacting fermion species. Consider for example the $SU(3)$ symmetric {\em attractive} Hubbard model involving three species of fermions on a bi-partite lattice whose Hamilton operator is given by
\begin{equation}
H = \sum_{\langle ij\rangle, a} -t (c^\dagger_{a,i} c_{a,j} + c^\dagger_{a,j} c_{a,i}) - 
V\sum_i \Big(N_i - \frac{3}{2}\Big)^2
\label{model2}
\end{equation}
where $a=1,2,3$ labels the three species. The operator $N_i = n_{1,i}+n_{2,i} + n_{3,i}$ is the total particle number at the site $i$. A straightforward extension of the discussion presented above solves the sign problem in this model. One begins by writing
\begin{equation}
\Big(N - \frac{3}{2}\Big)^2 = \frac{1}{2}\sum_{s,s'} s s'(n^{s}_1n^{s'}_2 + n^{s}_1n^{s'}_3 + n^{s}_2 n^{s'}_3)
\end{equation}
up to an overall constant. The partition function is then expanded as in (\ref{sse}) but now contains a product of three traces, one for each species. Each of these traces is written as a determinant and for the same reasons as discussed above, the sum over $[s]$ can be performed to get rid of diagonal terms in the matrices. Thus, only the off diagonal terms again contribute to each determinant and we finally obtain
\begin{equation}
Z \ = \ \sum_{[b]}\ \int\ [dt]\ (V)^k \Big\{\prod_{i=1,2,3} \mathrm{Det} (A_i([b,t]))\Big\}.
\end{equation}
Interestingly, one can show that the product of the three determinants is always positive. The reason is that while interactions can violate the reference configuration on each layer, the violations always come in pairs on two different layers. One can also add other interactions without introducing sign problems. For example the three body interaction of the type
\begin{equation}
H = \sum_i h_i n^{s_i}_{1,i}n^{s_i}_{2,i}n^{s_i}_{3,i}
\end{equation}
is allowed as long as $h_i \geq 0$ and $s_i$ is $+1$ on even sublattice and $-1$ on the odd sublattice.

The staggered reference configuration that plays an important role in the above solutions to the sign problem is related to the spontaneous symmetry breaking pattern that leads to the Mott insulating phase at large couplings. However, the same pattern forces all the above solvable models to remain at half filling. Thus, we still have not solved one of the most exciting sign problems that arises away from half filling. However, we believe that we may have taken an important step towards that solution. Note that we have managed to avoid the need for an additional layer of fermions to solve the sign problem. The next step would be to begin to dope the system and explore how sign problems creep in as we add particles (or holes) into the system. This may then give us further hints as to the origin of these more difficult sign problems and hence perhaps suggest solutions.

It is a pleasure to thank  H.~G.~Evertz, J.~Gubernatis, D.~Lee, R.~Scalettar, R.~Sugar, D.~Toussaint, S.~Zhang, and U.-J.~Wiese for helpful discussions and comments. This work was supported in part by the Department of Energy grants DE-FG02-05ER41368.

\bibliography{ref}
\end{document}